\documentclass{an}
\usepackage{graphicx}
\usepackage{times}
\usepackage{fancyhdr}
\newcommand{\lsim}{\raisebox{-0.6ex}{$\stackrel{{\displaystyle<}}{\sim}$}}

\renewcommand{\vec}[1]{\mbox{\boldmath$#1$}}

\begin{document}
\title{Oscillating  $\alpha^2$-dynamos and the reversal phenomenon  of the global geodynamo}
\author{A. Giesecke, G. R\"udiger, and D. Elstner}
\institute{Astrophysikalisches Institut Potsdam, An der Sternwarte 16, D-14482 Potsdam, Germany}
\date{Received; accepted; published online}

\abstract{A geodynamo-model based on an $\alpha$-effect which has been computed
under conditions suitable for the geodynamo is constructed. 
For a highly restricted class of radial $\alpha$-profiles the linear
$\alpha^2$-model exhibits oscillating solutions on a timescale given by the
turbulent diffusion time. 
The basic properties of the periodic solutions are presented and the influence
of the inner core size on the characteristics of the critical range that allows for
oscillating solutions is shown. 
Reversals are interpreted as half of such an oscillation.
They are rather seldom events because they can only occur if the $\alpha$-profile exists
long enough within the small critical range that allows for periodic solutions. 
Due to strong fluctuations on the convective timescale the probability of such a
reversal is very small. 
Finally, a simple non-linear mean-field model with reasonable input parameters based on
simulations of Giesecke et al. (2005) demonstrates the  plausibility
of the presented theory with  a long-time series of a (geo-)dynamo reversal sequence.  
}
\correspondence{agiesecke@aip.de}

\maketitle

\section{Introduction}
Paleomagnetic measurements show that the Earth's magnetic field exists for more than $10^9$ years with nearly the
same magnitude (Kono \& Tanaka 1995).
The process that is responsible for the production of this field is called the
{\em{geodynamo}} and essentially takes place in the fluid outer part of the
Earth's core.
The magnetic field is dominated by a dipole which -- as the most characteristic
feature --
from time to
time ``starts to oscillate'' and changes its polarity from one sign
to the other.
This phenomenon has been called {\em{reversals}} and typically lasts $10^3-10^4$
years (Bogue \& Merill 1992). 
The duration of such ``oscillating'' phase is extremely short compared with the time
between consecutive reversals. 
In fact
the average time between two reversals is about 50 times longer than
the duration of the reversal itself.
The appearance of the reversals
seems to be chaotic rather than periodic (Krause \& Schmidt 1988),
% in fact the
%paleomagnetic observations show that double reversals (which at least
%would cover a time span half of a full oscillating period) did never
%occur. 
and the probability of a reversal during a certain time span can be described by
a Poisson distribution.
On geological time scales ($\sim 10^7 - 10^8$ years) the average rate of
reversals changes (Merill et al. 1996) and there exist even very long periods were no reversal occurs -- so
called {\em{superchrons}}. 

Closely related to the reversal phenomenon are so called {\em{excursions}} a
kind of aborted reversals, where the polarity begins to change but, instead of
executing a full transition, the dipole returns to the original polarity.
Excursions occur about ten times more often than reversals.

Deviations from a perfectly axisymmetric field become manifest in the tilt of
the dipole axis with respect to the rotation axis (currently $11^{\circ}$)
and in the non-axisymmetric field components in terms of localized flux patches. 
Such field patterns in average exhibit a common directed drift motion, the so
called {\em{westward drift}} (see e.g. Bloxham \& Jackson 1989; Bloxham \& Jackson 1992).
Roughly simplifying the value of this westward drift amounts
approximately $0.3^{\circ}\!/\!{\mbox{year}}$.  
%
%\begin{itemize}
%\item 
%dipole only approximately parallel to the rotation axis. Currently this
%dipole-tilt is $11^{\circ}$ and the magnetic north-pol is located somewhere in
%north Canada. This location is not stationary ... behavior of the dipole axis
%on long timescales
%\item 
%secular variation
%\item 
%westward drift of flux patches
%\end{itemize}

Simulations of the 3D MHD-equations (that describe a
thermal/chemical driven turbulent flow of a conducting fluid in the
Earth's outer core and the magnetic field that is induced by this flow) have been able to
reproduce some of the observed features of the Earth's magnetic field
(Glatzmaier \& Roberts 1996; Kageyama \& Sato 1997; Christensen et al. 1998;
Kuang \& Bloxham 1999). 
Unfortunately such calculations are computational very expensive. The time periods that can be
covered are rather short compared to the time scales on which for example changes of the mean
reversal rate occur. In order to examine geodynamo-models in matters of the
statistics of the reversal phenomenon mean-field models remain indispensable.

A further unsolved issue is the influence of the small scales.
Global MHD simulations are restricted in the achievable parameter regime and in
the affordable spatial resolution. 
These limits prevent
from resolving the actual scales of the turbulence in the fluid outer core,
and for reasons of numerical stability unphysical large values for the viscous
losses have to be adopted. 
Therefore the smaller scales are artificially damped,
and properties and influence of the small scale turbulence remains unsure. 

Sarson \& Jones (1999) developed a  2.5D
model to examine the reversal mechanism in detail.
Their general picture is a ``large scale'' $\alpha\Omega$-dynamo mechanism,
where a strong zonal flow and meridional circulations are responsible for the dynamo action.
Reversals are induced by fluctuations of the meridional flow.  
The authors reclaim
that a opposite dipole polarity can evolve if the velocity remains long enough in a
regime that allows for oscillating magnetic fields.

Statistical properties in a one-dimensional
$\alpha\Omega$-dynamo model where reversals are triggered by a fluctuating
$\alpha$-effect have been examined by Hoyng et al. (2002).
Their model was able to predict some basic features of the Sint-800 data and although there have been some contradictions in
general their mean-field approach seems to be selfconsistent.

Other models, based on the absence of differential rotation (or any shear
flows) in the Earth's core, interpreted the geodynamo as an $\alpha^2$-dynamo
(Steenbeck \& Krause 1966).
It is known that the spherical $\alpha^2$-dynamo ``almost always'' possesses
stationary axisymmetric magnetic field solutions (for scalar $\alpha$-effect) or
longitudinally drifting non-axisymmetric solutions
(R\"udiger et al. 2003)
and therefore such models have difficulties to explain the reversal phenomenon. 
There are few exceptions of the  rule that $\alpha^2$-dynamos with scalar
$\alpha$-effect are stationary. 
%
%{\it
%\citet{1983PEPI...33..260O} retrieved rapid dipole field reversals in an
%$\alpha^2$-model where net helicity and hence the $\alpha$-effect change
%their sign due to random variations in energy release rates.
%%
%}

Fearn \& Rahman (2004) solved the Navier-Stokes equation and a mean-field
induction equation for an $\alpha^2$-dynamo model with a radial dependence of
the $\alpha$-effect given by $\alpha\propto\sin\pi(r-R_{\mathrm{in}})$.
This radial profile leads to a vanishing $\alpha$-effect
at the boundaries of the fluid outer core but the $\alpha$-effect does not
show any zero within the interior.
They obtained non-linear periodic
solutions if the $\alpha$ was larger than a certain critical
value. 
In contrast to the solutions of Sarson \& Jones (1999) their results were
strongly influenced by the non-linear back-reaction of the
Lorentz force on the flow which serves as a saturation mechanism for the
magnetic field.

Without considering any mean flow, Stefani \& Gerbeth (2003) found oscillating
$\alpha^2$-dynamos in case that the $\alpha$-effect (uniform in $\theta$) changes its sign in radius. 
Already
Soward (1974) with his quasi-linear approximation  for rotating convection in
layers with uniform density found that the $\alpha$-effect strongly varies with depth:
it is negative (positive) in the lower (upper) part of the convection layer --
well described by a radial sinus-function. Giesecke et al. (2005) have shown that
such profiles indeed result from numerical simulations of the convection in the
outer fluid core where the density stratification is very small (see Sect.~\ref{sec3}). 

In the present paper we shall show that radial $\sin$-profiles of the $\alpha$-effect lead to oscillating
$\alpha^2$-dynamos but already a slight deviation from this  profile provides
stationary modes. 
Combining the principle properties of the $\alpha$-effect from the calculations of
Giesecke et al. (2005) with a simple $\alpha^2$-dynamo, a mean-field model is
constructed that exhibits irregular reversals induced by a fluctuating
$\alpha$-effect. 
%
%%%%%%%%%%%%%%%%%%%%%%%%%%%%%%%%%%%%%%%%%%%%%%%%%%%%%%%%%%%%%%%%%%%%%%%%%%%%%%%%%%%%%%
%
%
%
\section{The equations}
Taking the induction equation 
\begin{equation}
\frac{\partial\vec{B}}{\partial t}=\nabla\!\times\!\bigg(\vec{u}\!\times\!\vec{B}\!-\!{\eta}\ \!\nabla\!\times\!\vec{B}\bigg)
\label{ind_eq}
\end{equation}
 and split magnetic field $\vec{B}$ and velocity $\vec{u}$ in a mean part,
 $\left<\vec{B}\right>$, $\left<\vec{u}\right>$ and a
fluctuating component $\vec{B}'$, $\vec{u}'$ the mean magnetic field
 $\left<\vec{B}\right>=\vec{B}-\vec{B'}$ is determined by
\begin{equation}
\frac{\partial \langle\vec{B}\rangle}{\partial t}=\nabla\times \bigg(\langle\vec{u}\rangle
\times
\langle\vec{B}\rangle+\vec{\mathcal{E}}-\eta\nabla\times\langle\vec{B}\rangle
\bigg)
\label{mf_ind_eq}
\end{equation}
with $\vec{\mathcal{E}}=\left<\vec{u}'\times\vec{B}'\right>$ as the mean
electromotive force (EMF) and $\eta$ the (molecular) magnetic diffusivity.
The components of the EMF are usually given by  
\begin{equation}
{\mathcal{E}}_i = 
\alpha_{ij}\langle{B}_j\rangle+\beta_{ijk}\partial_k\langle B_j\rangle.
\label{emf_eq}
\end{equation}
The tensor  $\alpha_{ij}$  correlates the EMF due to turbulent motions with the large-scale magnetic field, 
including the effects of
anisotropy.
In the simplest case the tensor $\beta_{ijk}$
is related to the turbulent diffusivity by $\beta_{ijk}=\eta_{\mathrm{T}} 
\epsilon_{ijk}$ which is the case that we shall discuss here.
In the following all mean flows $\langle\vec{u}\rangle$ are neglected and we
end up with a mean-field induction equation that describes an $\alpha^2$-dynamo:
\begin{equation}
\frac{\partial \langle\vec{B}\rangle}{\partial t}=\nabla\times \bigg(\alpha\langle\vec{B}\rangle-\eta_{\mathrm{T}}\nabla\times\langle\vec{B}\rangle\bigg),
\label{alphameanfield_eq}
\end{equation}
where $\eta_{\mathrm{T}}\gg\eta$ is implied.
Equation~(\ref{alphameanfield_eq}) together with a prescribed $\alpha$-effect
that depends on the radius $r$ and the latitude angle $\theta$ is solved
numerically using an explicit finite difference scheme in two dimensions in
spherical coordinates. The standard resolution is $64 \times 64$ grid points in both
radial and latitudinal directions. 
A perfect conductor is assumed to exist at the inner core boundary which is
justified because of the absence of turbulent motions
in the solid inner core so that the diffusivity is significant smaller 
inside the solid core than inside the fluid outer core. 
However, a finitely conducting inner core affects the behavior
of the magnetic field as it has been shown e.g. by Hollerbach \& Jones (1993).
At the
outer boundary a vacuum is simulated by increasing the magnetic dissipation by a factor of 10.
The details of the numerical realizations of these
boundary conditions are described by R\"udiger et al. (2003).
To estimate general properties of the linear mean-field model at first a quenching
mechanism is abandoned to avoid the complicated questions that are associated with the
non-linearities. 
Only for the long time simulations in Sect.~\ref{4} an equilibration
mechanism is used to prevent the field from growing to infinity.
%
%
%
%
%%%%%%%%%%%%%%%%%%%%%%%%%%%%%%%%%%%%%%%%%%%%%%%%%%%%%%%%%%%%%%%%%%%%%%%%%%%%%%%%%
%
%
%
%
\section{Geodynamo $\alpha$-effect}\label{sec3}
\subsection{General properties}
\begin{figure}[h!]
\hspace*{0.5cm}
\vspace*{0.5cm}
\includegraphics[width=8cm,height=5cm]{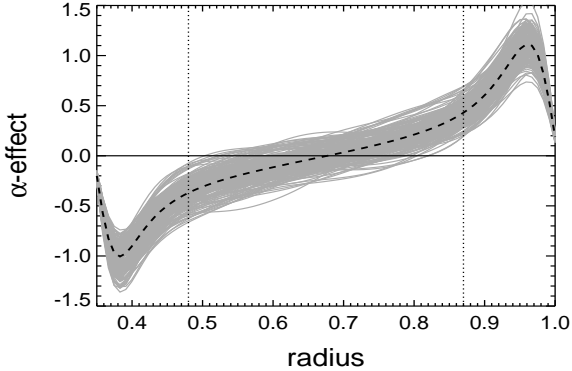}
\caption{ Radial dependence of the $\alpha$-effect for ${\rm \Lambda} =
  \displaystyle\frac{\vec{B}^2}{2\mu\rho\Omega\eta}=1$. The dashed thick line denotes the
  time average and the grey lines show fluctuations in time.}
\label{alpha}
\end{figure}

Fig.~\ref{alpha} shows a typical radial dependence of the $\alpha$-effect computed
from local simulations of weak stratified and fast rotating magnetoconvection
by Giesecke et al. (2005).  
The $\alpha$-effect on the northern hemisphere shows a maximum (minimum) close to the upper
(lower) boundary and the cross-over takes place exactly in the middle of the
convective unstable layer. 
The radial profile is almost perfectly antisymmetric with respect to the
middle of the layer so that the net $\alpha$-effect (integrated over $r$)
approximately vanishes.

The dotted lines divide the domain in three
parts. 
In the outer zones the exact determination of the $\alpha$-effect is 
difficult because close to the boundaries strong gradients
of the magnetic field require the exact knowledge of the turbulent diffusivity
$\eta_{\mathrm{T}}$ for an calculation of the $\alpha$-effect from Eq.~(\ref{emf_eq}).
In the central part -- between $r=0.48$ and $r=0.87$ -- the field
gradients are negligible and thus the result should be more reliable. 
But even if the presented profile comes with some uncertainties, the qualitative
overall behavior can quite well be described by $\alpha\sim \sin r$, where the
argument of the $\sin$ must be chosen in a way that the $\alpha$-effect
disappears at the inner and the outer core boundary.
If a density stratification is included in the simulations, the
$\alpha$-effect cross-over moves more and more towards the bottom of the
box (see R\"udiger \& Hollerbach 2004 their Fig.~4.23). 
\subsection{Oscillating $\alpha^2$-dynamos}
The general properties of the above presented $\alpha$-effect are used as an
input for a global axisymmetric $\alpha^2$-dynamo.
The model
is a spherical shell with the inner radius $R_{\rm in}$ and the outer radius
$R_{\rm out}$. 
For the present date Earth the radius of the solid inner core is given by
$R_{\mathrm{in}}=1222\,\mathrm{km}$ and the radius of the fluid outer core is
given by $R_{\mathrm{out}}=3480\,\mathrm{km}$. 
In all simulations the radius of the outer core is scaled to $1$ and to maintain
the correct ratio, the
radius of the inner core is scaled to $0.35$. 
In the following the $\alpha$-tensor is  antisymmetric with respect to
the equator ($\sim \cos\theta$).
The ``standard profile'' of the $\alpha$-effect is given by
\begin{equation}
\alpha(r,\theta)=\alpha_0\cos\theta\sin\left(2\pi\frac{r-R_{\mathrm{in}}}{R_{\mathrm{out}}-R_{\mathrm{in}}}\right).
\label{alpha_profile}
\end{equation}
This equation is slightly modified to vary the amplitude in the upper (lower)
half of the sphere and the zero-crossing of the $\alpha$-effect.
We start with a strict $\sin$-profile in radius of the $\alpha$-effect, i.e. with a
cross-over $R_0$ in the middle  between $R_{\rm in}=0.35$ and $R_{\rm out}=1$ and equal
amplitudes. If supercritical the field grows exponentially and the resulting dynamo oscillates.

\begin{figure}[htb]
\hspace*{0.6cm}
\includegraphics[width=8cm,height=5cm]{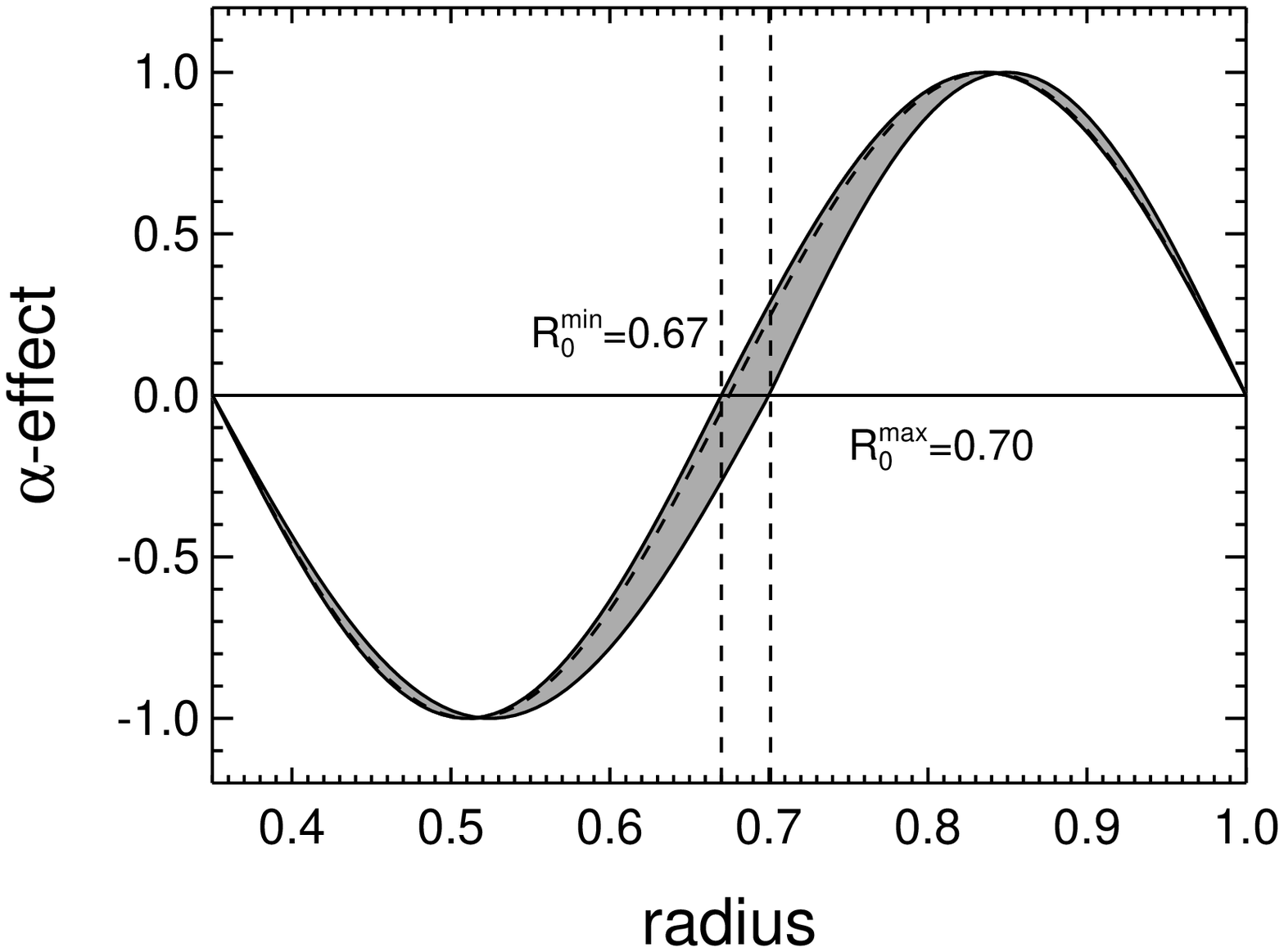}\\
\hspace*{0.6cm}
\includegraphics[width=8cm,height=5cm]{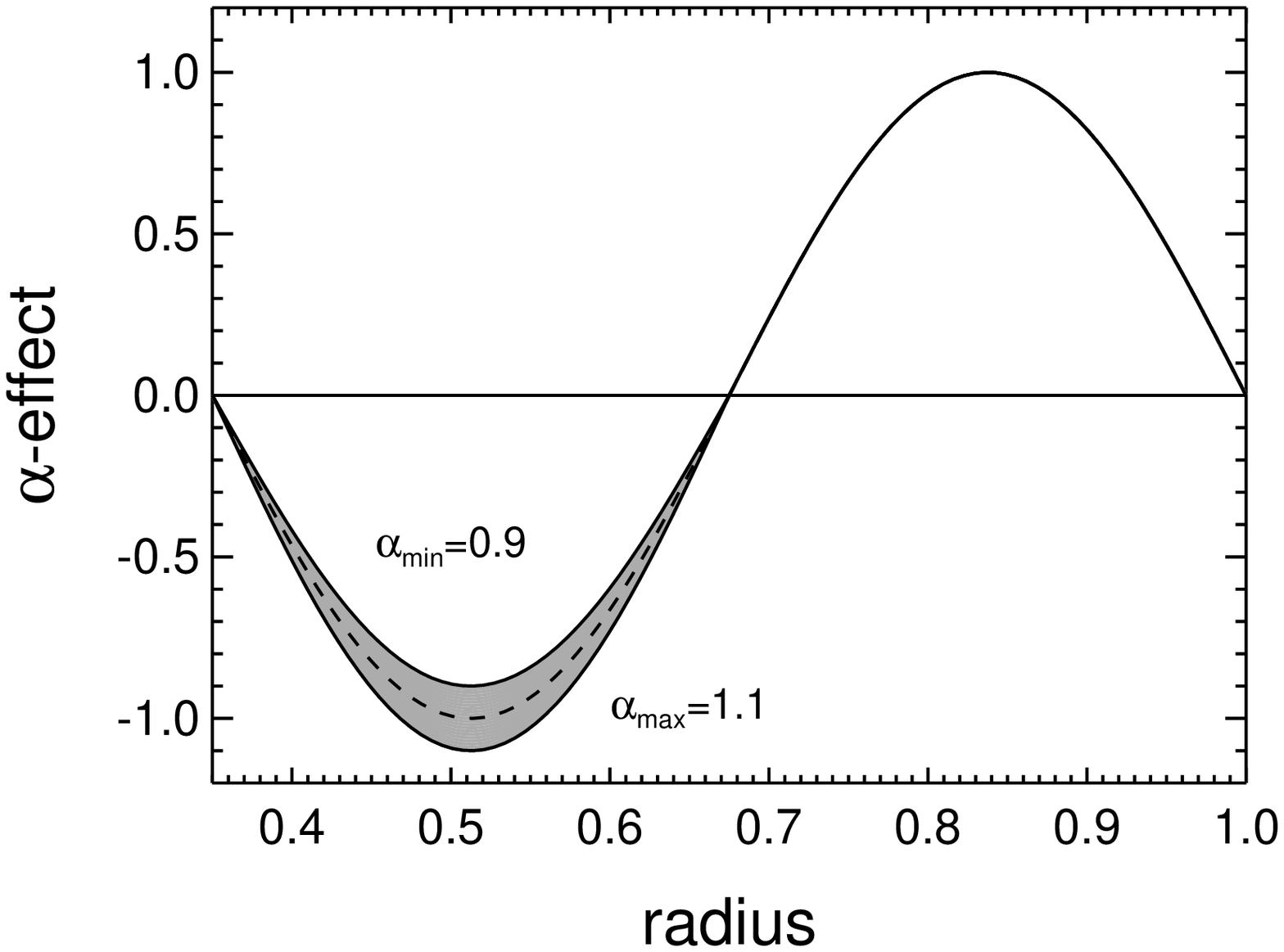}
\vspace*{0.5cm}
\caption{Critical domains for the radial profile of the $\alpha$-effect that
  lead to oscillating solutions. Top: critical interval for the
  location of the zero. Bottom: critical interval for the magnitude of
  the lower amplitude} 
\label{crit_zero}
\end{figure}

%
%Obviously the oscillation frequency is determined by the diffusion time $R_{\rm
%out}^2/\eta$. 
%
%
The condition to
$R_0$ for the excitation of the oscillation of $\alpha^2$-dynamos is very
strict. 
The oscillations only exist for $R_0\approx 0.67...0.70$. 
%(if too long oscillations are excluded, see Fig.~\ref{f1}). 
%
If the radial
$\alpha$-profile does not lie in the narrow area indicated in Fig.~\ref{crit_zero}
(top) then
the dynamo does not oscillate. 
The periodic solution due to the strict radial $\sin$-profile of the
$\alpha$-effect can also be suppressed by a variation of one of the amplitudes.
Numerical experiments were made by multiplication of the lower part of the
sin-profile with a factor $A$. Oscillating solutions are found with $\delta
A\simeq 0.1$ (see bottom of Fig.~\ref{crit_zero}). Of course, a similar result would hold for the upper part of
the radial $\sin$-profile but the general property of the model  is now known: The oscillating solution only
exists if the deviations of the actual $\alpha$-effect profile from the
profile given by Eq.~(\ref{alpha_profile}) are very small.

Due to fluctuations a simple radial $\sin$-profile of the
$\alpha$-effect is rather seldom. 
A possible oscillation of the dynamo only happens if the profile of the
$\alpha$-effect lasts long enough within the critical range that allows for
periodical solutions. 
The minimum time which must be
covered by the (critical) radial $\alpha$-profile in order to excite (half of) an
oscillation has been estimated from the simulations and is given approximately by
\begin{equation}
t_{\mathrm{min}}\approx0.3\cdot \tau_{\rm diff}
\label{tmin_eq}
\end{equation} 
with the diffusion time $\tau_{\mathrm{diff}}$ defined by
\begin{equation}
\tau_{\mathrm{diff}}=\frac{R_{\mathrm{out}}^2}{\eta}.
\label{diff_time_eq}
\end{equation}
If this minimum time is not reached by the radial profile of the $\alpha$-effect the
stationary solution would not start to become oscillating. 

In all simulations the time the dipole needs for the transition from one
polarity to the other is of the order of $\tau_{\mathrm{diff}}$.
If we assume $\eta\simeq 2\cdot 10^4\ {\rm cm}^2/{\rm s}$ (for molten iron under conditions in the fluid
outer core) we retrieve $\tau_{\mathrm{diff}}\sim 10^5$ years. 
To keep the model consistent with the observed duration of a reversal ($10^4$
years) we have to assume a
turbulence-induced enhancement of the magnetic diffusivity by one order of
magnitude: 
\begin{equation}
\eta_{\mathrm{T}}\approx 10\cdot \eta. 
\label{turdiff}
\end{equation}
With $\eta_{\mathrm{T}} \approx 20 \mathrm{m}^2\!/\!\mathrm{s}$ the diffusion
time reduces to $10^4$ yrs, the typical duration of a reversal. 
The ratio $\eta_{\mathrm{T}}/\eta\approx 10$ seems not to be totally devious and was also the result of a rough
estimation of Giesecke et al. (2005). 
A comparable value has been presented by Hoyng et al. (2002) who
independently determined a turbulent diffusion time of $10000-15000$ years from
the analyzation of 
autocorrelation functions from the Sint-800 observations. 
However, it should be kept in mind, that
this are very rough estimations. 
\subsection{Field pattern of a reversal}
\begin{figure*}[htb!]
\hspace*{-1.cm}
\includegraphics[width=9cm]{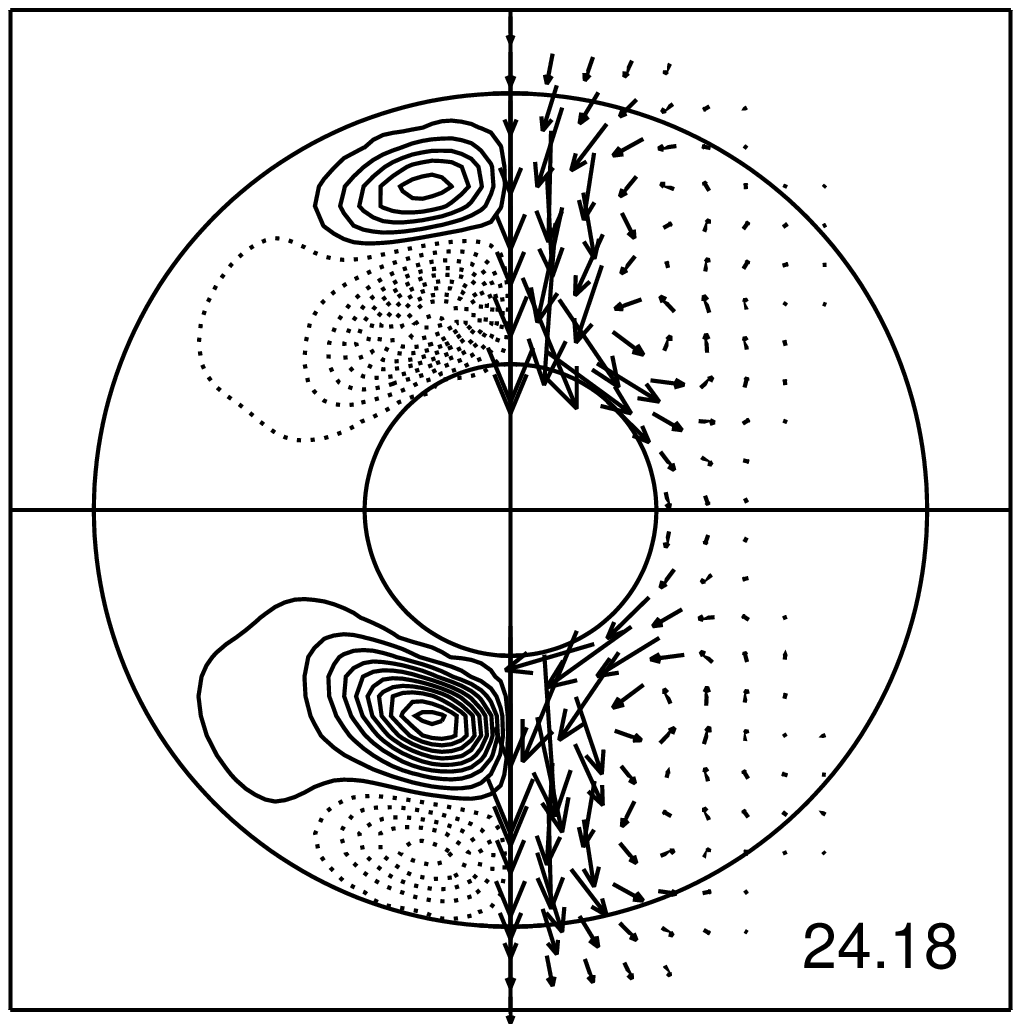}
\hspace*{-4cm}
\nolinebreak[4]
\includegraphics[width=9cm]{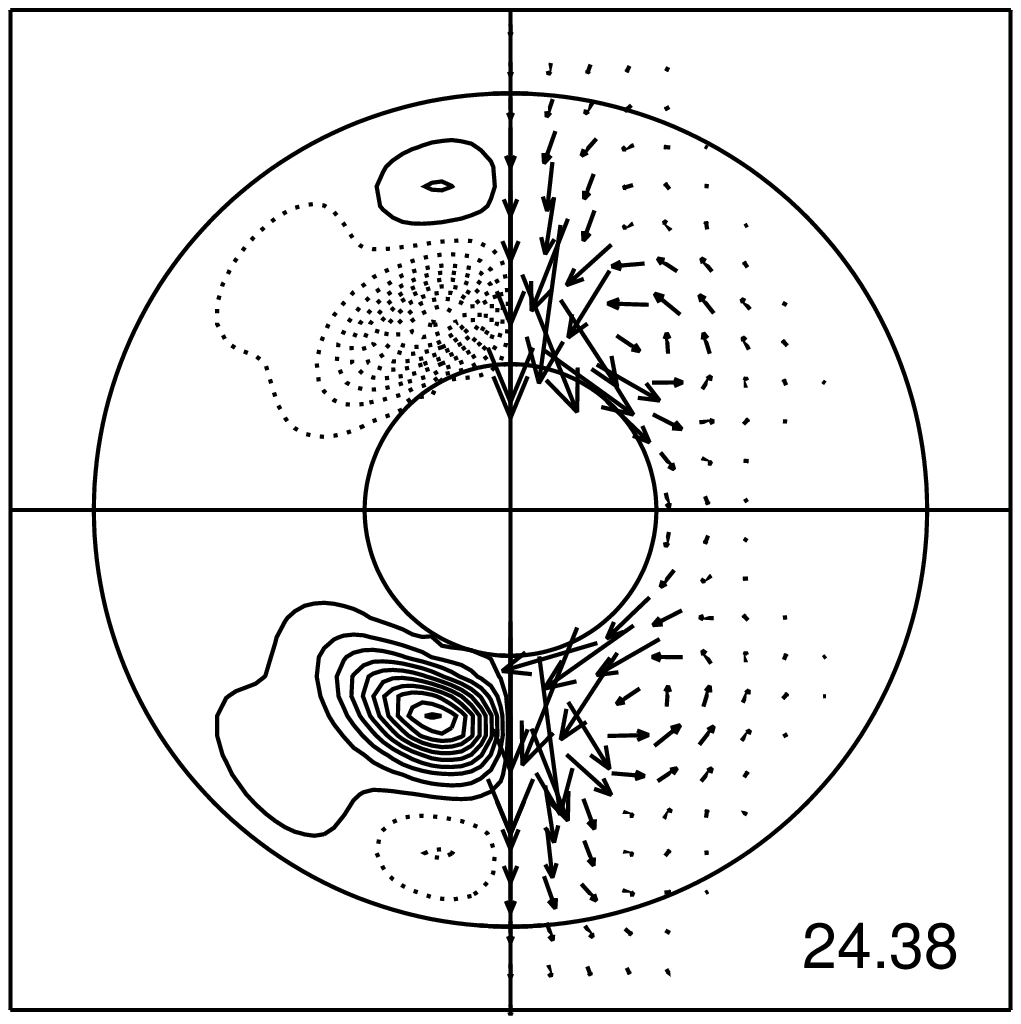}
\nolinebreak[4]
\hspace*{-4cm}
\includegraphics[width=9cm]{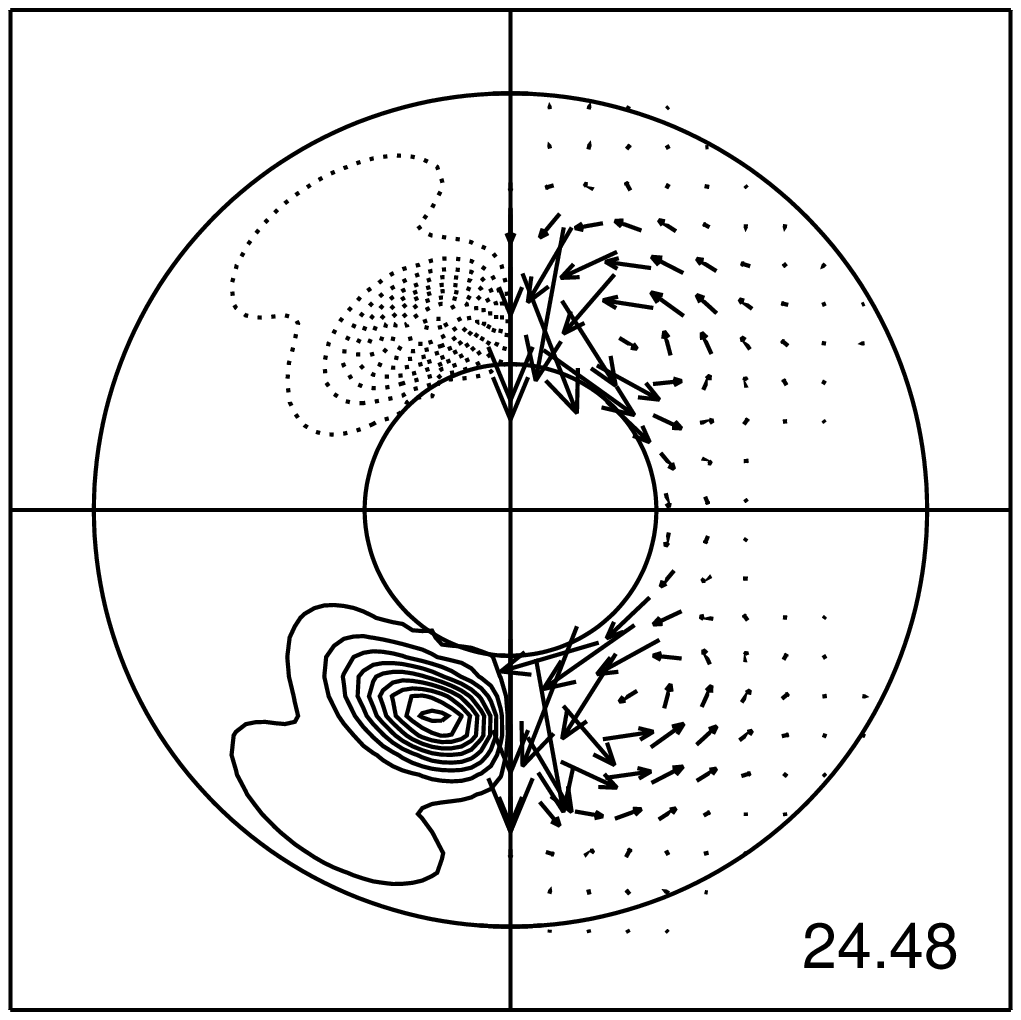}\\[-1.3cm]
\hspace*{-1.cm}
\includegraphics[width=9cm]{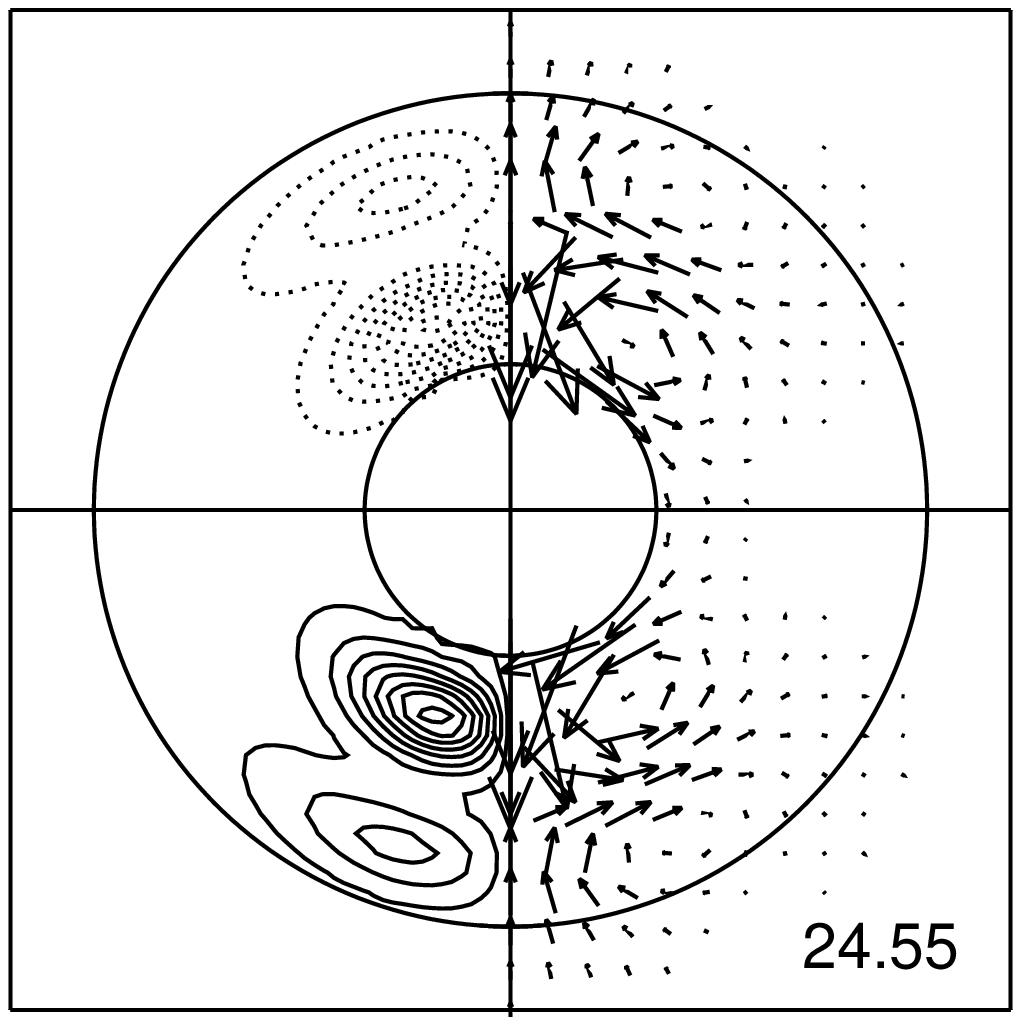}
\nolinebreak[4]
\hspace*{-4cm}
\nolinebreak[4]
\includegraphics[width=9cm]{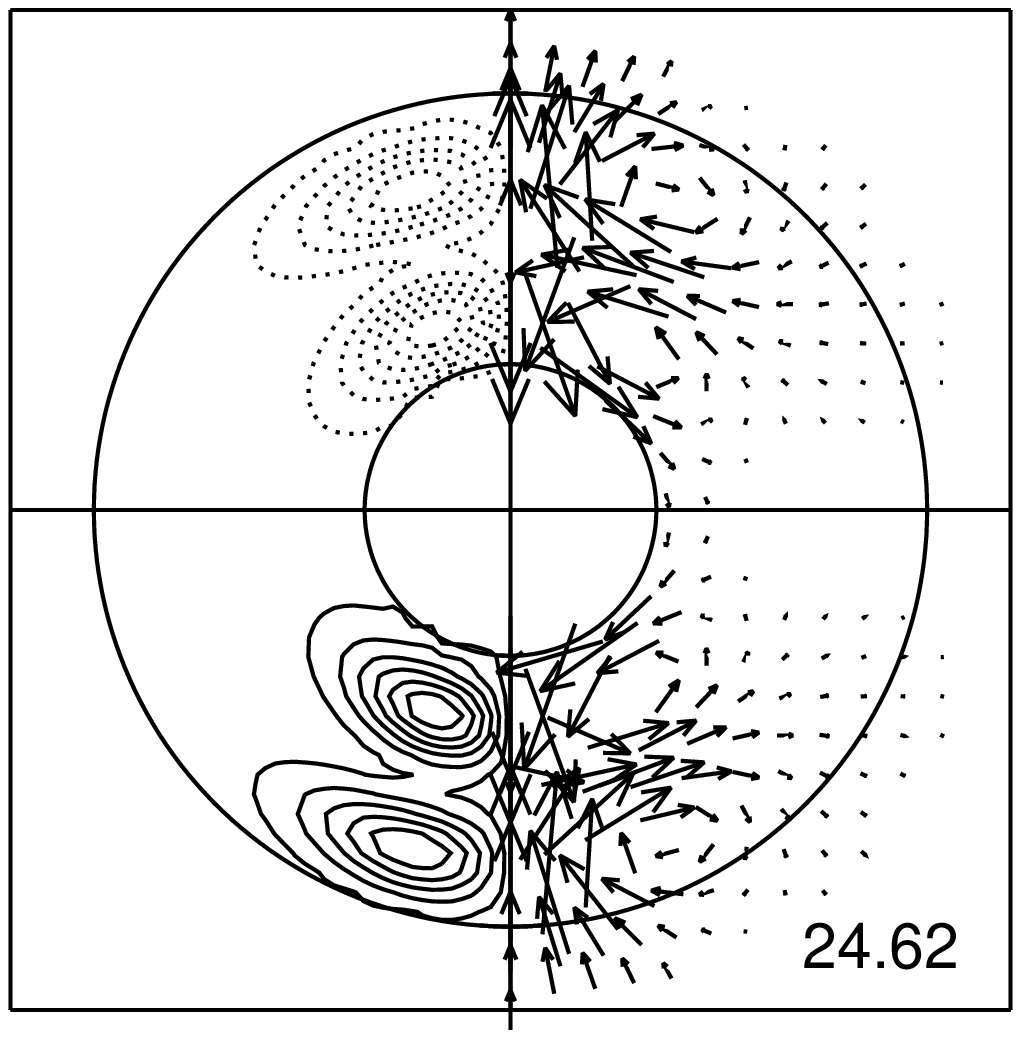}
\nolinebreak[4]
\hspace*{-4cm}
\includegraphics[width=9cm]{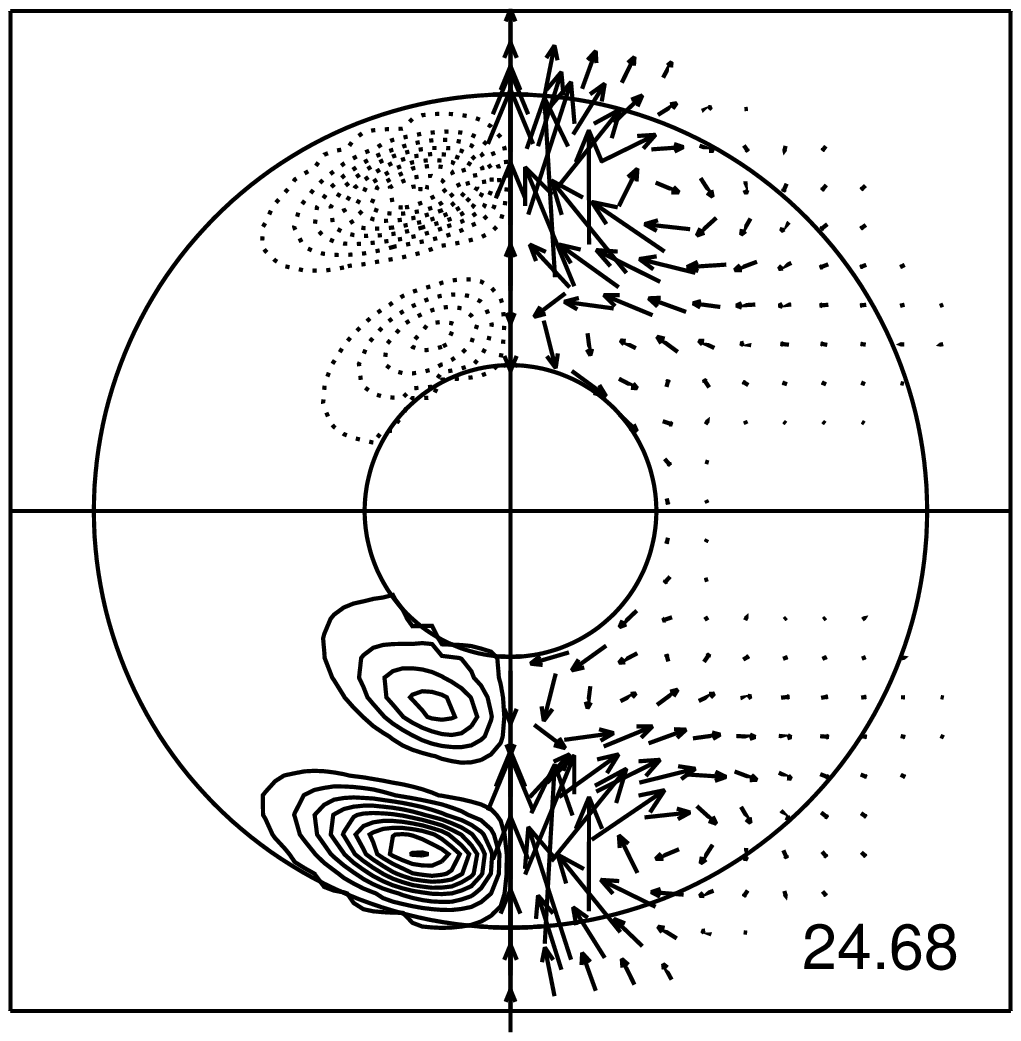}\\[-1.3cm]
\hspace*{-1.cm}
\includegraphics[width=9cm]{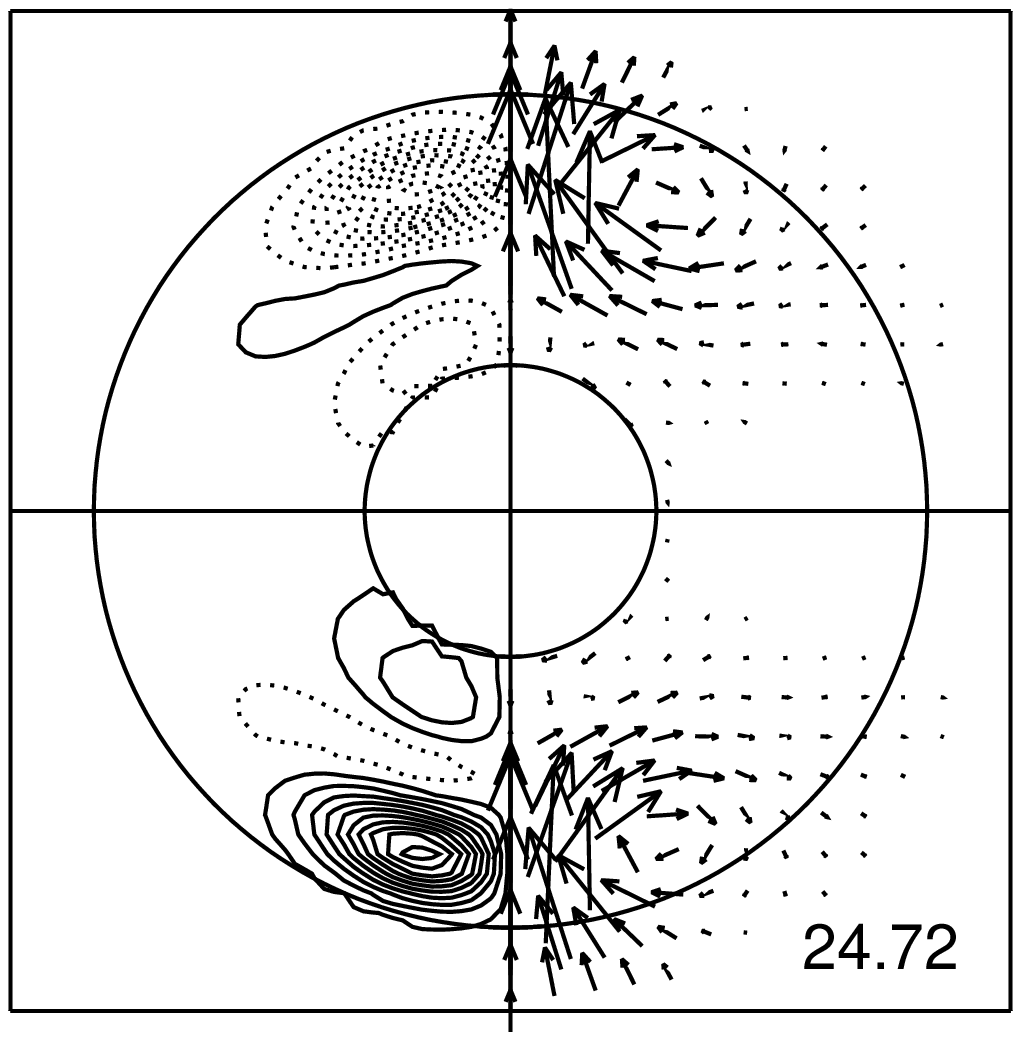}
\hspace*{-4cm}
\nolinebreak[4]
\includegraphics[width=9cm]{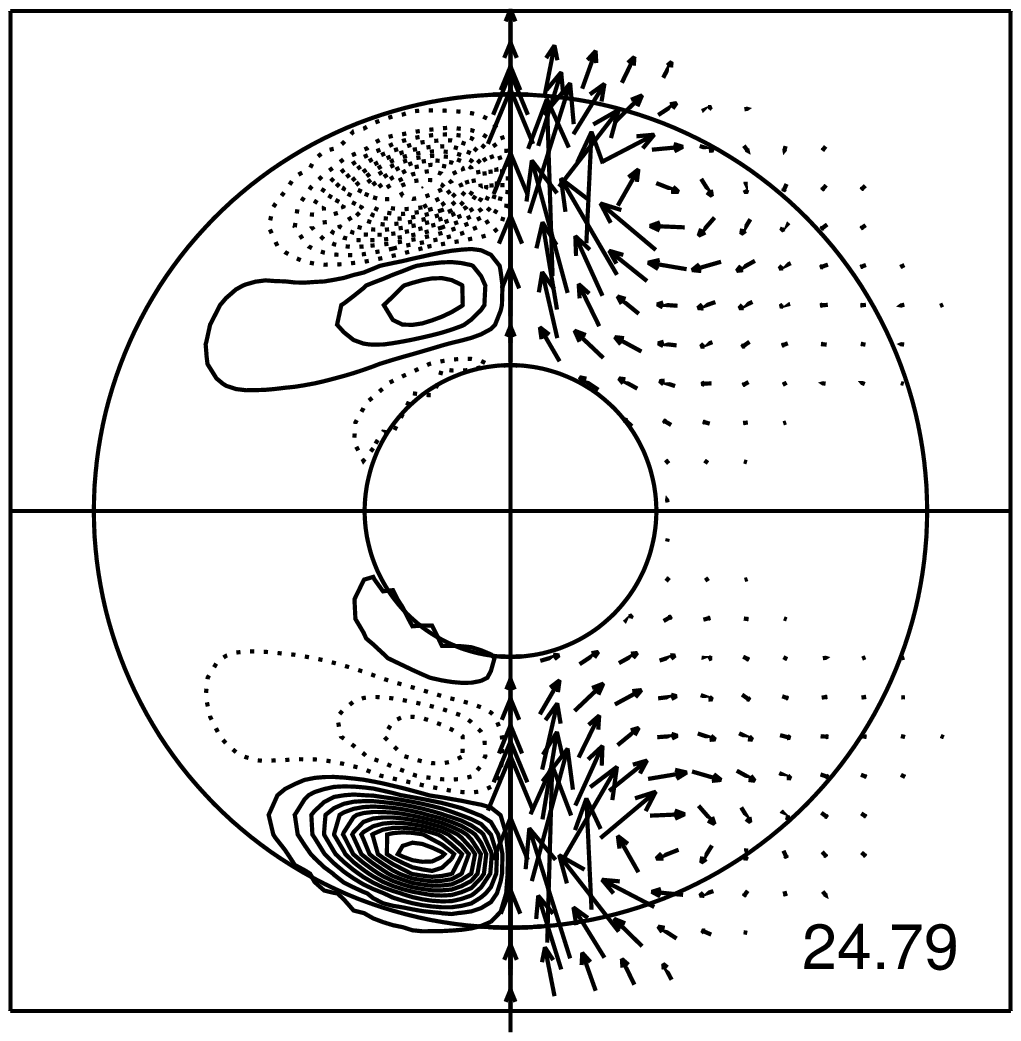}
\nolinebreak[4]
\hspace*{-4cm}
\includegraphics[width=9cm]{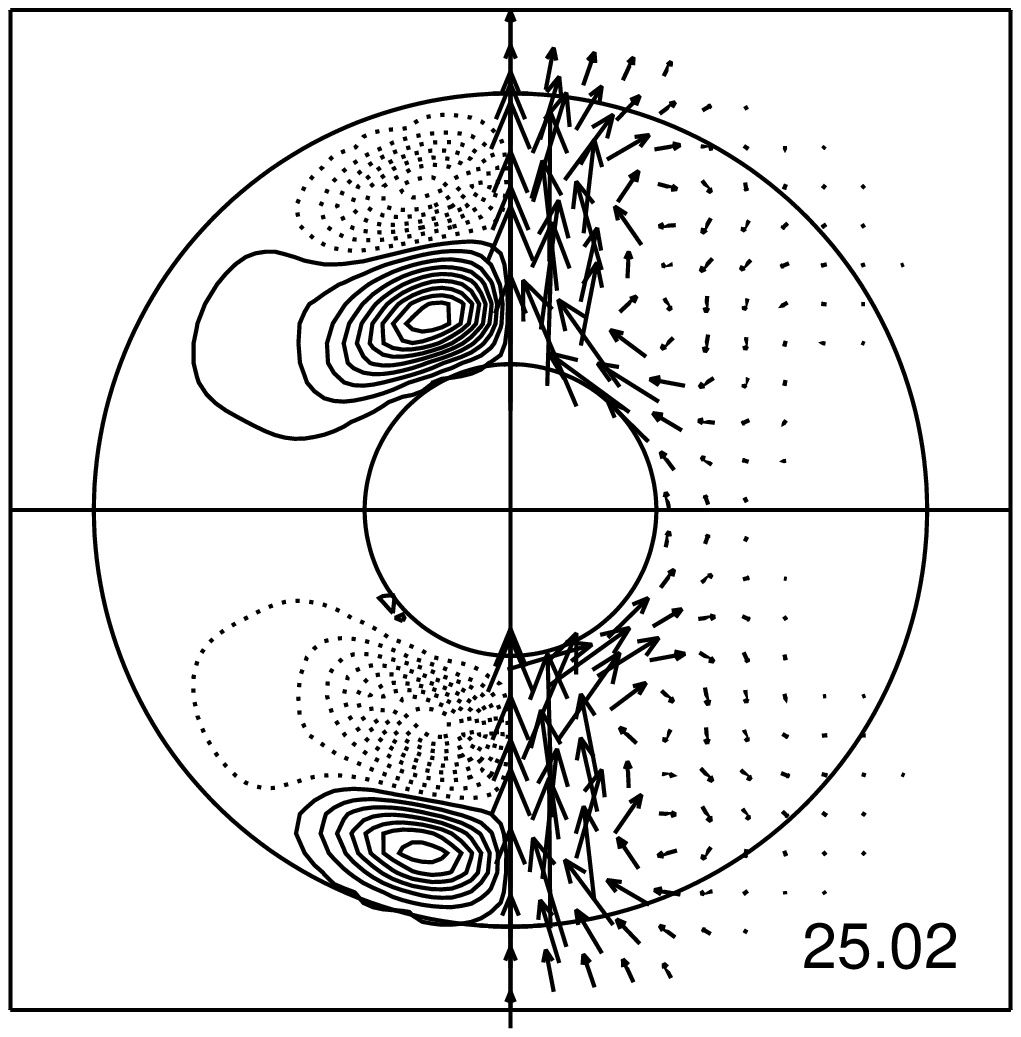}
\caption{The magnetic field configuration of the  cyclic solution with a strict radial $\sin$-profile of the $\alpha$-effect. The time in units of the diffusion time
 is printed in the lower right corner
  of each snapshot.} 
\label{contour_rev}
\end{figure*}

Figure~\ref{contour_rev} shows the temporal behavior of the magnetic field projected on a
meridional plane during one oscillation. 
The left-hand side of each panel shows the isolines of the toroidal field component where solid (dashed)
lines denote field directions clockwise (counterclockwise). 
The arrows on the right hand side represent the poloidal field component. The
length of the arrows is scaled with the field length.
The magnetic field is concentrated near the rotation axis which is a
consequence of the latitudinal 
$\cos\theta$-dependence of the $\alpha$-effect. The
solution is of dipolar parity.
Regarding the toroidal component in the northern hemisphere the considered reversal cycle
starts close to the rotation axis with a clockwise oriented toroidal
field in
the upper half of the  outer core and a field of opposite sign starts in the
lower part of the outer core.
The outer part weakens (1,2) and is replaced by
a growing toroidal field of opposite sign (2,3,4).
Between the two belts of equally counterclockwise (clockwise) oriented components 
that determine the appearance of the magnetic field in the northern (southern)
hemisphere in the middle of the reversal sequence (5,6) a new opposite directed field appears (7) and
pushes away the lower toroidal component (8).
The reversal is completed in panel (9) when the lower part of the sphere is
 completely filled out by this new reversed oriented field.
Note that the poloidal field component already shows the reversed polarity in
part (6) of Fig.~\ref{contour_rev} and only undergoes minor changes in the
remainder of the sequence.
%
%
%
%
%%%%%%%%%%%%%%%%%%%%%%%%%%%%%%%%%%%%%%%%%%%%%%%%%%%%%%%%%%%%%%%%%%%%%%%%%%%%%%
%
%
%
\subsection{Oscillation period and critical dynamo number}
\begin{figure}[htb!]
\includegraphics[width=8cm]{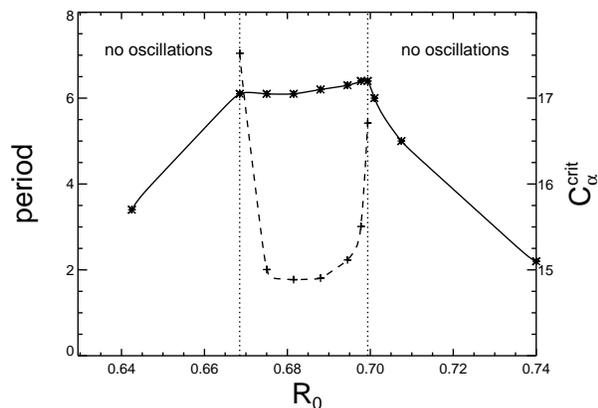}
\caption{The characteristics of the linear $\alpha^2$-dynamo model  in
dependence of the location of the cross-over point $R_0$ in the radial $\alpha$-profile.}
\label{periode_calpha}
\end{figure}

Figure~\ref{periode_calpha} shows the oscillation period
(dashed line, in units of $\tau_{\mathrm{diff}}$) and the
critical dynamo number (solid line)  
\begin{equation}
C_{\alpha}^{\mathrm{crit}}=\frac{\alpha^{\mathrm{crit}}R_{\mathrm{out}}}{\eta_{\mathrm{T}}}
\end{equation}
in dependence of the location of the zero of the $\alpha$-effect ($R_0$).
$C_{\alpha}^{\mathrm{crit}}$ determines the minimum amplitude of the $\alpha$-effect at which
dynamo-action occurs.
The vertical dotted lines indicate the transition between oscillating and
stationary solutions.
The critical dynamo number is always larger for the oscillating solutions. 
This is not a surprising result.
The oscillating solutions have smaller scales than the stationary solutions so
that the Ohmic losses are larger for the oscillating solutions. The immediate
consequence is that for given amplitude  of $\alpha$ the oscillating solutions
are less nonlinear than the non-oscillating solutions.
In the center of the
critical interval the oscillation-period is nearly constant for $R_0$ and
increases strongly if the zero of the $\alpha$-effect is close to the upper or
lower boundary that separates the oscillating solutions from the stationary states.
The values of the
critical dynamo number for the symmetric
and antisymmetric (with respect to the equator) axisymmetric modes (S0, A0) and for the first
non-axisymmetric modes (A1, S1) are specified in Table~\ref{crit_c_tab}.
Since the dominant part of the Earth's magnetic field is a dipole the A0 mode
is of profound interest.
Note that the A0-mode (leads to a dipole solution) and the S0-mode (leads
to a quadrupole solution) coincide within our numerical accuracy.
\begin{table}[htb]
\caption{Critical dynamo number for different basic dynamo modes for $\alpha$
  taken from Eq.~(\ref{alpha_profile}) ($R_0=0.675$)}
\label{crit_c_tab}
\begin{tabular}{ccccc}\hline
&{\bf{A0}} & {\bf{A1}} & {\bf{S0}} & {\bf{S1}}\\
\hline
\\
${\mathbf{{\it{C}}_{\alpha}^{\mathrm{crit}}}}$& 17.04 & 17.67 & 17.04 & 17.67 \\
\\
\hline
\end{tabular}
\end{table}
%
%
%
%
%%%%%%%%%%%%%%%%%%%%%%%%%%%%%%%%%%%%%%%%%%%%%%%%%%%%%%%%%%%%%%%%%%%%%%%%%%%%%%%%%%%%%%%%%%%%%
%
%
\subsection{Influence of the inner core size}
\begin{figure}[h]
\hspace*{-5cm}
\includegraphics[width=18cm,height=8cm,clip=TRUE]{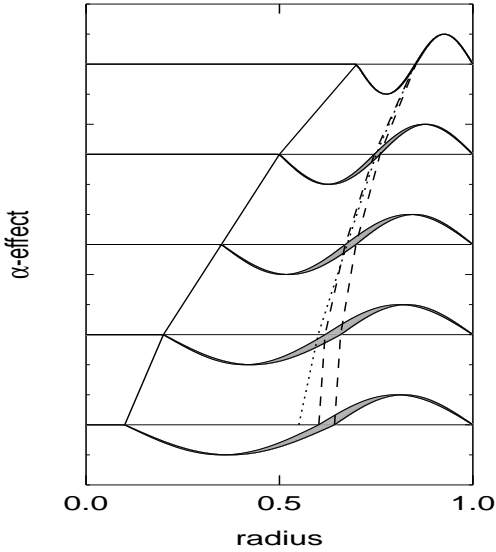}
\caption{Critical $\alpha$-profiles for oscillating solutions for different
  sizes of the inner core. From top to bottom:
  $R_{\mathrm{in}}=0.7,0.5,0.35,0.2,0.1$. The dashed lines confine the
  critical interval and the dotted line denotes the center of the fluid outer core.} 
\label{gap}
\end{figure}

%
%
%%%%%%%%%%%%%%%%%%%%%%%%%%%%%%%%%%%%%%%%%%%%%%%%%%%%%%%%%%%%%%%%%%%%%%%%%%%%%%%%%%%
%
%
The solid inner core of the Earth is growing on geological time scales
(hundreds of millions of years). 
Due to the specific thermodynamic conditions in
the Earth's interior the liquid iron first freezes out at the center of the
Earth because the melting point of iron decreases faster with increasing
pressure than the temperature increases towards the center. 
The appearance of
a solid inner core few billion years ago resulted in important changes of
the physical conditions and processes that dominate the flow in the fluid core.
It is obvious that the
size of the solid inner core also should have a strong influence on the oscillations of the
$\alpha^2$-model that has been presented above.
We restrict our examination to the geometric effects that arise from different
sizes of the inner core and do not consider the changes in the
turbulence that are associated e.g. with the emerging compositional convection.
Figure~\ref{gap} shows the critical profiles of the $\alpha$-effect that lead to
oscillating solutions for different ratios of $R_{\mathrm{out}}/R_{\mathrm{in}}$.
For increasing size of the inner core -- indicated by the solid vertical line
on the left side -- the critical interval becomes smaller (indicated by the
dashed line). At the
same time the center of this interval moves closer to the center of the fluid outer
core (denoted by the dotted line). 
For $R_{\mathrm{in}}=0.1$ (bottom curve) the critical
cross-over-points of the $\alpha$-profiles are clearly located in the upper
half of the fluid outer core, whereas for $R_{\mathrm{in}}=0.7$ (top curve) the critical
profile is nearly a perfect $\sin$-profile as given by Eq.~(\ref{alpha_profile})
where the zero is located exactly in the middle of the fluid outer core and
the width of the critical interval has become very small. 

Although the critical interval becomes
smaller with increasing size of the inner core -- making it more difficult to
excite an oscillation -- one can speculate that the
overall probability for a reversal in case of a fluctuating $\alpha$-profile
might be approximately constant (or at least $>0$) for all sizes since the center of the interval
moves towards the middle of the fluid outer core which is the preferred
location for the zero cross-over of the $\alpha$-effect (see Fig.~\ref{alpha}). 
\subsection{Non-axisymmetric modes}
The $\alpha$-effect is a nontrivial
tensor if the rotation is fast: $\alpha_{zz}\to 0$ for $\Omega \to \infty$
(in cylindrical ccordinates, see Moffatt 1970; R\"udiger 1978; Busse \& Miin 1979) where
$\alpha_{zz}$ refers to the $\alpha$-effect in cylindrical
coordinates and $z$ denotes the direction parallel to the rotation axis. 
The tensorial structure of the $\alpha$-effect is now taken into account, i.e. relation the $\alpha_{zz}=0$ is used in the dynamo equation.

\begin{figure}[h]
\includegraphics[width=8cm]{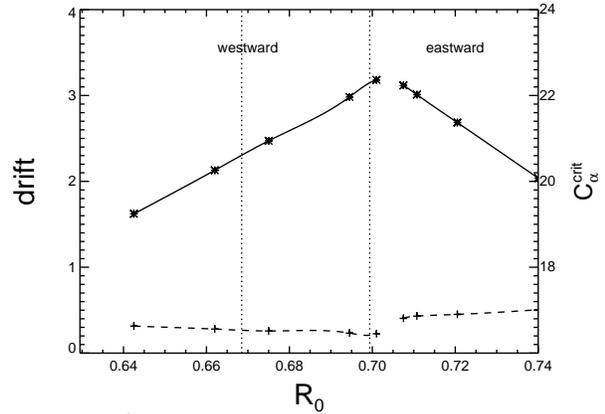}
\caption{$C_{\alpha}^{\mathrm{crit}}$  of the non-axisymmetric modes for the case $\alpha_{zz}=0$. The corresponding drift periods are given by the dashed line.} 
\label{drift_ps}
\end{figure}
Figure~\ref{drift_ps} shows $C_{\alpha}^{\mathrm{crit}}$ (solid line) and the
drift period (dashed line) for the lowest non-axisymmetric modes. Again the results for the symmetric mode (S1) and the antisymmetric mode
(A1) coincide in both quantities.
The dotted vertical lines denote the critical interval for oscillating
solutions for the A0- respective S0-mode for the (scalar) isotropic $\alpha$-tensor as
described in the previous section.

For the zero of the $\alpha$-effect below $R_0\approx
0.705$ we obtain a westward drift. Above $R_0\approx 0.71$ both modes show an
eastward directed drift motion.
The characteristic drift time
scale is approximately $(0.20...0.30)\!\cdot\!\tau_{\mathrm{diff}}$ for the
westward drifting modes and $(0.40...0.50)\!\cdot\!\tau_{\mathrm{diff}}$ for the
eastward drifting modes.
Note that the ratio of the drift timescale ($0.2\tau_{\mathrm{diff}}$) to
reversal timescale ($\tau_{\mathrm{diff}}$) obtained from the simulations
coincides with the same ratio observed for the Earth where the timescale of the westward
drift is $\sim 2000$ years and the duration of a reversal $\sim 10000$ years.
However, these timescales
are estimations with a large uncertainty (e.g. data for the reversal time reach from 100
to several 10000 years).

The critical
dynamo numbers for the axisymmetric modes (S0, A0) are  larger
than $C_{\mathrm{\alpha}}^{\mathrm{crit}}$ of the A1/S1-modes so that the solution would be dominated by these non-axisymmetric modes. The axisymmetric modes with the higher eigenvalues are oscillating.  We know that this constellation is
 in contradiction to the observations of the Earth's magnetic field
which is dominated by a stationary  dipole-part. The non-axisymmetric modes only lead
to smaller contributions that are manifested in the dipole tilt and the
drifting field patterns. The interaction of the nonaxisymmetric modes (for anisotropic $\alpha$) and the oscillating modes (for isotropic $\alpha$ with cross-overs) is still an open question.
%

%
%
%
%%%%%%%%%%%%%%%%%%%%%%%%%%%%%%%%%%%%%%%%%%%%%%%%%%%%%%%%%%%%%%%%%%%%%%%%%%%%%%%%%%%%%
%
%
%
\section{Irregular reversals induced by a fluctuating $\mathbf{\alpha}$-effect}
\label{4}
In the following the complications that appear from the
non-axisymmetric modes are ignored. The effects of the full non-trivial
$\alpha$-tensor will be treated in a subsequent paper.

\begin{figure}[htb]
\includegraphics[width=9cm,height=6cm]{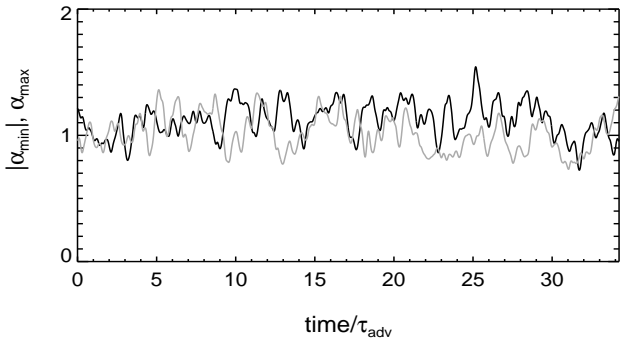}
\caption{Fluctuations of the maximum and the minimum of the $\alpha$-effect in
dependence of time (in units of the turnover-time $\tau_{\mathrm{adv}}$) taken from the results
shown in Fig.~\ref{alpha}.} 
\label{fluc_alpha}
\end{figure}

It is known from numerous calculations that the $\alpha$-coefficients are
rather noisy quantities (e.g. Ossendrijver et al. 2001; Giesecke et al. 2005).
Figure~\ref{fluc_alpha} shows the maximum (minimum) of the
$\alpha$-effect in dependence of the time in units of the turnover time or
advective timescale taken from the simulations of Giesecke et al. (2005)
\begin{equation}
\tau_{\mathrm{adv}}=\frac{R_{\mathrm{out}}-R_{\mathrm{in}}}{u'}
\end{equation}
(where $u'$ is the turbulent rms-velocity).

The amplitude of the $\alpha$-effect in the outer part of the shell is
slightly larger than in the inner part and the strength of the fluctuations
amounts approximately 10\% of the average.
\begin{figure*}[htb]
\vspace*{-4cm}
\includegraphics[width=18cm,clip=TRUE]{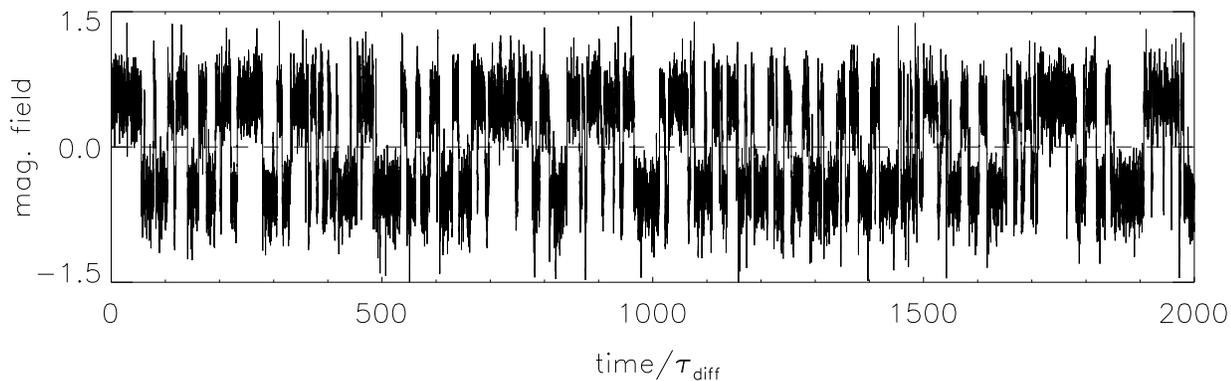}
\\[-3.5cm]
\caption{Reversals of the magnetic field. 173 events have been identified as
  reversals which leads to a mean polarity life-time of
  approximately 11 diffusion times.} 
\label{rev_field}
\end{figure*}
It is obvious that the timescale of the fluctuations is
given by $\tau_{\mathrm{adv}}$.
If we assume a typical value for the Earth's fluid outer core:
$u'\approx 5\cdot 10^{-4}\mathrm{m}/\mathrm{s}$ the timescale of the
fluctuations $\tau_{\mathrm{adv}}$ is related to the diffusive timescale $\tau_{\mathrm{diff}}$ by 
\begin{equation}
\tau_{\mathrm{adv}}\approx (0.01...0.02)\cdot\tau_{\mathrm{diff}}
\label{adv_timescale}
\end{equation}
where $\tau_{\mathrm{diff}}$ is estimated by Eq~(\ref{turdiff}).
Thus the duration which is necessary for the $\alpha$-effect to stay in a
critical state -- given by Eq.~(\ref{tmin_eq}) -- is about $15... 30$ times longer than
the time scale on which the $\alpha$-effect fluctuates. This indicates that a
reversal clearly must be a very seldom event.
It is also typical for the presented theory of the reversal phenomenon of the
geodynamo that practically never a realization of the $\alpha$-profile may
exist so long that a complete oscillation can happen.

For the long time calculations we adopt an isotropic $\alpha$-effect given by
Eq.~(\ref{alpha_profile}) and add fluctuations for the magnitude and for the location of the zero. 
For simplicity we assume equal averages for the upper and lower amplitude
which both vary independently. 
The fluctuations
are described by a Gaussian distribution with a standard deviation $\sigma$ of 10\% of
the average. 
The fluctuations of the zero cross-over are slightly larger. We adopt an
average value of $R_0=0.675$ (the middle of the fluid layer) with  a standard deviation of $\sigma=\pm0.100$. 
According to Eq.~(\ref{adv_timescale}) the actual values for the fluctuating
quantities are updated each $0.01 \tau_{\mathrm{diff}}$. 

The dynamo number is given by $C_{\alpha}$ = 20. This is clearly overcritical
and therefore the non-linearities introduced by a local quenching
function for the $\alpha$-effect given by 
\begin{equation}
\alpha(\vec{B})=\alpha_0\frac{1}{1+{\vec{B}^2}}
\label{quench_eq}
\end{equation}
might also have some influence on the solution. Indeed test calculations show
that increasing $C_{\alpha}$ -- corresponding to a stronger driven and thus
more non-linear dynamo -- reduces the probability of a reversal.

\begin{figure}[htb]
\hspace*{0.6cm}
\includegraphics[width=8cm,height=4cm]{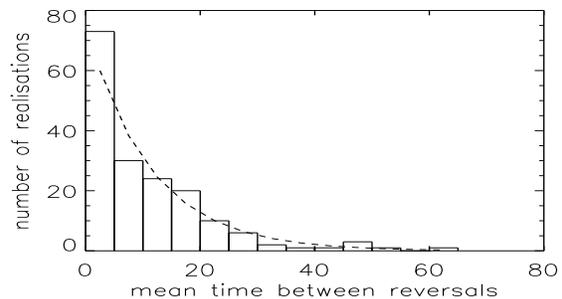}
\vspace*{0.5cm}
\caption{Number of realizations of life-times of a single polarity state.} 
\label{lifetime}
\end{figure}

Figure~\ref{rev_field} shows the time dependence of the radial magnetic field at
some point in the spherical shell for a long time calculations that spans 2000
diffusion times (with the actual time scaling this corresponds to $20$ million years). 
The magnetic field reverses irregularly. Both polarity states occur with
nearly the same probability and field strength. 
The distribution of time-periods between each reversal can be described by an
exponential function ${\rm e}^{-\Delta t/t}$, thus the reversals are independent
randomly occurring events (see Fig.~\ref{lifetime}).
The number of short life-times between consecutive reversals is overestimated
because it is difficult to distinguish between excursions and
reversals. 
Within the presented 2D-simulations the only possibility to characterize a
reversal is a sign change of a field component. 
Other properties, e.g. an slight increase of the tilt of the dipole axis
followed by an immediate decrease back to the original state as an indication
for an excursion, are intrinsically not available.  
Here we filtered out all events where the field changes its sign only
``slightly'' and for a very short time ($\lsim\,0.5\,\tau_{\mathrm{diff}}$). This means
that an original dipole-state recovers very fast after the magnetic
field just touches the zero in Fig.~\ref{rev_field}. 
For the chosen set of parameters the mean reversal rate is about a factor of 5
higher than the rate observed for the present date Earth.
Short-time test-calculations also have been performed with slightly different values for $\eta_{\mathrm{T}}/\eta$. 
This is the crucial
relation that specifies the $\alpha$-effect fluctuation timescale with respect to the turbulent diffusion
time.
The choice of this timescale fixes the time span for which the $\alpha$-effect remains in a
certain state and therefore determines the probability of a
reversal. 
Decreasing $\eta_{\mathrm{T}}/\eta$ increases the (turbulent) diffusion time
which results in a decrease of the relation
$\tau_{\mathrm{adv}}/\tau_{\mathrm{diff}}$ (assuming that
$\tau_{\mathrm{adv}}$ is fixed for the Earth's case). 
Therefore the timescale of the
fluctuations of the $\alpha$-effect is shorter with respect to the diffusive
timescale and the probability of a reversal decreases. 
The results show that already a slightly reduced value for
$\tau_{\mathrm{adv}}/\tau_{\mathrm{diff}}$ significantly increases the mean
time between consecutive reversals.

%
%
%
%
%
%
%%%%%%%%%%%%%%%%%%%%%%%%%%%%%%%%%%%%%%%%%%%%%%%%%%%%%%%%%%%%%%%%%%%%%%%%%%%%%%
%
%
%
%
%
\section{Discussion}
We have shown that an $\alpha^2$-dynamo is able to exhibit periodic solutions
where the dipole polarity changes on a timescale of the order of the diffusion
time $\tau_{\mathrm{diff}}$.
A crucial contribution for the existence of an oscillating $\alpha^2$-dynamo is an
$\alpha$-effect that strongly depends on the radial coordinate $r$. 
The most
important property of such $\alpha$-effect is a sign change within the field
producing domain and a certain symmetry between positive and negative magnitude. 
Only a highly
restricted class of (radial) $\alpha$-profiles leads to  oscillating
solutions. 
It turned out that the radial profiles of the $\alpha$-effect that have been
calculated by Giesecke et al. (2005) under conditions that are suitable for
the Earth's fluid core provide the essential characteristics
that are necessary to construct such oscillating dynamos. 

Long time simulations over 20 million years retrieve solutions 
which show irregular reversals. 
The model is very sensitive to the choice of parameters that describe the
statistical properties.
In principle it should easily be possible to
adopt values for the fluctuating quantities ($C_{\alpha}$, $\sigma$ and
$\tau_{\mathrm{adv}}/\tau_{\mathrm{diff}}$) in a way that would result in a
time series which exhibit the
basic properties of the geodynamo.
The most important characteristics are the mean reversal rate, the distribution of
polarity states, the distribution of the mean time between reversals and the ratio
between number of excursions to number of reversals.
An exact adjustment of the model at this stage is beyond the purpose of this
work because further effects that affect the
behavior of the field as well have not been considered. 
We neglected the influence of a finitely conducting inner core. 
Test calculations have shown that this does not prevent the solutions from
oscillating, only some properties like the positions of the critical intervals
might be slightly changed.
Possible changes in the behavior of the convection (changes in amplitude and
strength of the fluctuations) have been ignored just as the (possible)
existence of large-scale flows.

An unsolved issue is the inclusion of the non-trivial components of the
$\alpha$-effect. Currently it is not possible to retrieve the characteristics
of the non-axisymmetric parts of the Earth's magnetic field. The Ansatz
$\alpha_{zz}=0$ -- based on theoretical reasons --  resulted in a
dynamo that is dominated by the non-axisymmetric modes (and the axisymmetric
modes, dipole and quadrupole are strongly suppressed). 
A resulting realistic solution should be described by a
combination of different modes as it seems to be the case for the geodynamo.  
Indeed, observations of the magnetic fields of other planets or moons in
the Solar system show, that various manifestations of field configurations are possible. 
The nearly perfect dipole field of Saturn or the non-axisymmetric dominated fields from
Uranus and Neptune are some extraordinary examples.
%
%
%%%%%%%%%%%%%%%%%%%%%%%%%%%%%%%%%%%%%%%%%%%%%%%%%%%%%%%%%%%%%%%%%%%%%%%%%%%%%%%%%%%%%%%%%%%
%
%
%
%
%
\acknowledgements
{This work was supported by the DFG SPP ``Erdmagnetische Variationen''.}

%\bibliographystyle{astron}

%\bibliography{references}

\begin{thebibliography}{}
\bibitem{}
Bloxham, J., Jackson, A.: 1989,
JGR 94, 15753

\bibitem{}
Bloxham, J., Jackson, A.: 1992,
JGR 97, 19537

\bibitem{}
Bogue, S.~W., Merill, R.~T.: 1992,
AREPS {20}, 181

\bibitem{}
Busse, F.~H., Miin, S.~W.: 1979,
GAFD {14}, 167

\bibitem{}
Christensen, U., Olson, P., Glatzmaier, G.~A.: 1998,
GRL {25}, 1565

\bibitem{}
Fearn, D.~R., Rahman, M.~M.: 2004,
GAFD {98}, 385

\bibitem{}
Giesecke, A., Ziegler, U., R\"udiger, G.: 2005,
PEPI, in press

\bibitem{}
Glatzmaier, G.~A., Roberts, P.~H.: 1996,
Physica D {97}, 81

\bibitem{}
Hollerbach, R., Jones, C.~A.: 1993,
PEPI {75}, 317

\bibitem{}
Hoyng, P., Schmitt, D., Ossendrijver, M.~A.~J.~H.: 2002,
PEPI {130}, 143

\bibitem{}
Kageyama, A., Sato, T.: 1997,
PhRvE {55}, 4617

\bibitem{}
Kono, M., Tanaka, H.: 1995, 
 in: T. {Yukutake} (ed.), {\em The Earth's Central Part: Is Structure and Dynamics},
 Terrapub, Tokyo, Japan

\bibitem{}
Krause, F., Schmidt, H.-J.: 1988,
PEPI {52}, 23

\bibitem{}
Kuang, W., Bloxham, J.: 1999,
JCP {153}, 51

\bibitem{}
Merill, R.~T., McElhinny, M.~W., McFadden, P.~L.: 1996,
 {\em The Magnetic Field of the Earth, Paleomagnetism, the Core and the
  Deep Mantle},
 Academic, London

\bibitem{}
Moffatt, H.~K.: 1970,
JFM {44}, 705

\bibitem{}
Ossendrijver, M., Stix, M., Brandenburg, A.: 2001,
A\&A {376}, 713

\bibitem{}
R\" udiger, G.: 1978,
AN {299}, 217

\bibitem{}
R\"udiger, G., Elstner, D., Ossendrijver, M.: 2003,
A\&A {406}, 15

\bibitem{}
R\"udiger, G., Hollerbach, R.: 2004,
 {\em The Magnetic Universe - Geophysical and Astrophysical Dynamo
  Theory},
 Wiley-VCH Verlag Berlin

\bibitem{}
Sarson, G.~R., Jones, C.~A.: 1999,
PEPI {111}, 3

\bibitem{}
Soward, A.~M.: 1974,
Royal Society of London Philosophical Transactions Series A
  {275}, 611

\bibitem{}
  Steenbeck, M., Krause, F.: 1966,
Zeitschrift f. Naturforschung A {21}, 1285

\bibitem{}
Stefani, F., Gerbeth, G.: 2003,
PhRvE {67}, 027302

\end{thebibliography}
\end{document}